\documentclass[letterpaper, 10pt, conference]{IEEEconf}      

\usepackage{amsmath} 

\usepackage{hyperref}
\usepackage{physics}
\usepackage{siunitx}
\usepackage{amssymb}
\usepackage{xcolor}
\usepackage{caption}
\usepackage{subcaption}
\usepackage{mathtools}
\usepackage{graphicx}
\usepackage{textcomp}

\newtheorem{remark}{\textbf{Remark}}

\newcommand{\tran}{^{\mkern-1.5mu\mathsf{T}}}

\title{\LARGE \bf
Experimental Platform for Boundary Control of Mechanical Frenkel-Kontorova Model 
}

\author{Loi Do, Krištof Pučejdl, and Zdeněk Hurák
\thanks{This work was supported by the Grant Agency of the Czech Technical University in Prague, grant No. SGS22/166/OHK3/3T/13 and by the Czech Science Foundation
(GACR) under contract No. 21-07321S.}
\thanks{Loi Do, Krištof Pučejdl, and Zdeněk Hurák are with Faculty of Electrical Engineering, Czech Technical University in Prague
        {\tt\small \{doloi, kristof.pucejdl, hurak\}@fel.cvut.cz}}%
}

\begin{document}

\maketitle
\thispagestyle{empty}
\pagestyle{empty}


\begin{abstract}
In this paper, we present a laboratory mechatronic platform for experimental verification and demonstration of various dynamical and control system phenomena exhibited by the Frenkel-Kontorova (FK) model -- a spatially discretized version of the sine-Gordon equation. 
The platform consists of an array of torsionally coupled pendulums pivoting around a single shaft that can be controlled through the motors at boundaries while all angles are read electronically. 
We first introduce and describe the platform, providing details of its mechatronic design and software architecture.
All the files are freely shared with the research community under an open-source license through a public repository. 
The motivation for this sharing is to help reproducibility or research -- the platform can be useful for others as a testbed for control algorithms for this class of dynamical systems, for instance, distributed control or control of flexible structures. 
In the second part of the paper, we then showcase the platform using two control problem formulations tailored to the FK model -- we discuss the practical motivation for studying these problems, propose (some) methods for solving them and demonstrate the functionality using experiments. 
In particular, the first control formulation deals with non-collocated control/regulation of a single pendulum through one boundary of the array in the presence of a disturbance sent from the other boundary; the second control problem consists in synchronizing pendulums' angular speeds. Some other motivated problems can certainly be formulated, solved, and demonstrated using the proposed platform. 


\end{abstract}

\begin{keywords} 
Frenkel-Kontorova model, sine-Gordon equation, multi-pendulum mechatronic platform, boundary control
\end{keywords}


\section*{Supplementary Materials}
All supplementary materials, such as assembly instructions, source codes, and bills of materials, are available at \href{https://github.com/aa4cc/The-Frenkel-Kontorova-laboratory-model}{github.com/aa4cc/The-Frenkel-Kontorova-laboratory-model}.
The repository also includes additional multimedia files, for instance, videos from experiments.


\section{Introduction}
The \textit{Frenkel-Kontorova} (FK) model, first presented in~\cite{frenkel_theory_1938} within a study of crystal dislocation, describes a motion of a one-dimensional chain of harmonically coupled identical particles in a spatially periodic potential field.
Subsequently, the model was found to describe many other physical systems, for instance, \textit{Josephson junction} arrays, mechanical properties of an open-state of the DNA, and systems in solid-state physics; see~\cite{cuevas-maraver_sine-gordon_2014} for further details.
The reason behind the FK model's universality is arguably the fact that it represents a discretized version of the celebrated \textit{sine-Gordon} (sG) equation.

In this paper, we present a mechatronic platform realizing the FK model as an array of torsionally coupled pendulums pivoting around a single shaft and show two novel boundary-control formulations defined on the FK model.
The idea of the mechanical platform realizing the FK model is not new. 
However, as we document in the next section, our platform differs from other works in several aspects.
The platform can be useful for educational demonstrations or as a testbed for research in many areas of automatic control, such as distributed control of coupled oscillators, control of underactuated systems, or control of lump-modeled flexible structures.


\section{Related Work}
Many works studied the FK model as an array of pendulums, primary in a connection to the sG equation.
In~\cite{scott_nonlinear_1969}, the author presented and highlighted a pedagogical benefit of how the mechanical FK model can be used to demonstrate the non-linear wave phenomena of the sG equation.
The model was constructed as a series of spring-connected nails soldered into a brass base and supported horizontally on a taut wire attached at both ends to a frame.

Later, an improved construction of the FK model was presented in~\cite{thakur_driven_2007} and~\cite{cuevas-maraver_discrete_2009}.
The authors studied a particular solution of the sG equation, so-called \textit{Intrinsic Localized Modes} (ILMs), also referred to as \textit{discrete breathers}.
Unlike~\cite{scott_nonlinear_1969}, the array's frame was additionally subjected to a harmonic driving in a horizontal direction.
By applying suitable open-loop control of the driving's amplitude and frequency, the array forms a stable, localized, or even moving ILM.
A similar mechanical setup was developed in~\cite{ikeda_intrinsic_2015}, where two disc-type pendulums with an elastic cord as a spring were used to confirm the theoretical analysis of ILM's behavior under imperfections in mechanical parameters.

In a series of papers~\cite{fradkov_control_2008},~\cite{fradkov_multipendulum_2012}, the authors presented another mechatronic multi-pendulum setup based on the FK model.
Compared to~\cite{thakur_driven_2007, cuevas-maraver_discrete_2009}, this setup was directly designed as a testbed for feedback control tasks.
The chain can be controlled by motors attached at the boundaries, and, as the theoretical works by the same authors deal with distributed control, each pendulum can be individually controlled by an electromagnetic actuator.
An optical encoder in each module provides measurements of the pendulum's angle.

The laboratory model, which we present in this paper, is mechanically comparable to~\cite{fradkov_control_2008, fradkov_multipendulum_2012}.
Similarly, every pendulum in the chain has its angle measured by an encoder, and the model can be controlled by motors attached to its boundaries.
In contrast, the pendulums in our platform are all attached to a single base, so the model is compact and easily scalable. 
Additionally, as the platform is open-source, it is relatively simple to reproduce and can also be modified to particular needs of the scientific community.


\section{Experimental Platform Description}
\begin{figure}[!t]
        \centering
        \includegraphics[width=8.4cm]{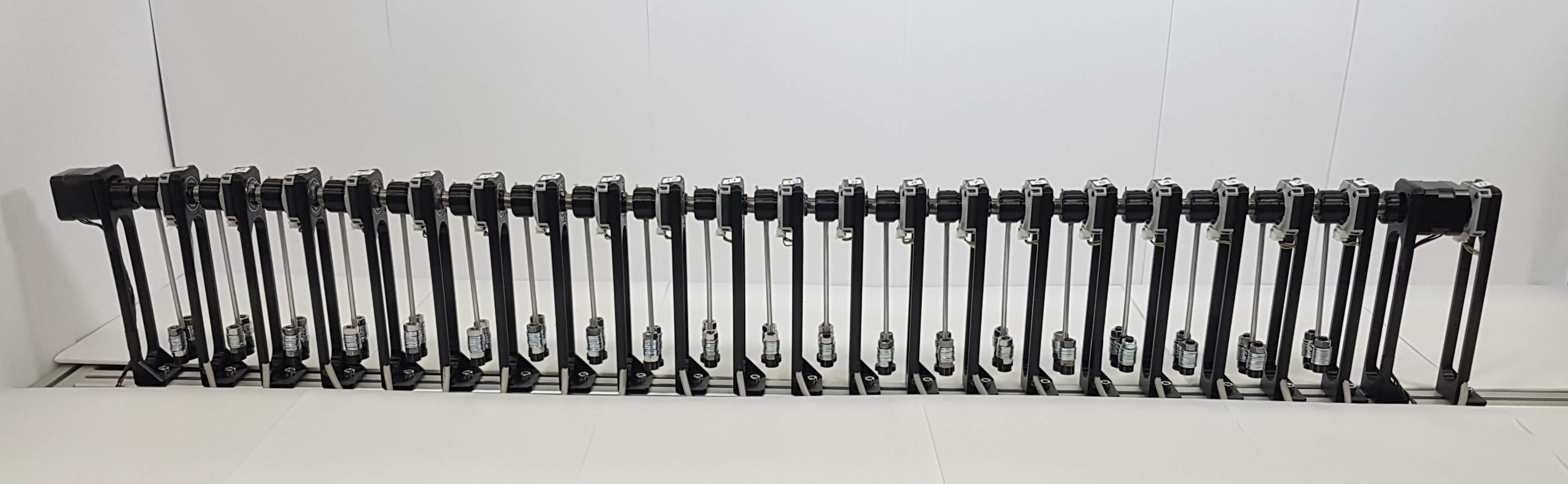}
        \caption{A photo of the platform with 20 pendulum modules and two motors attached at the boundaries}\label{fig:temp_FK_lab_model_complete}
\end{figure}

\begin{figure}[!t]
        \centering
        \includegraphics[width=8cm]{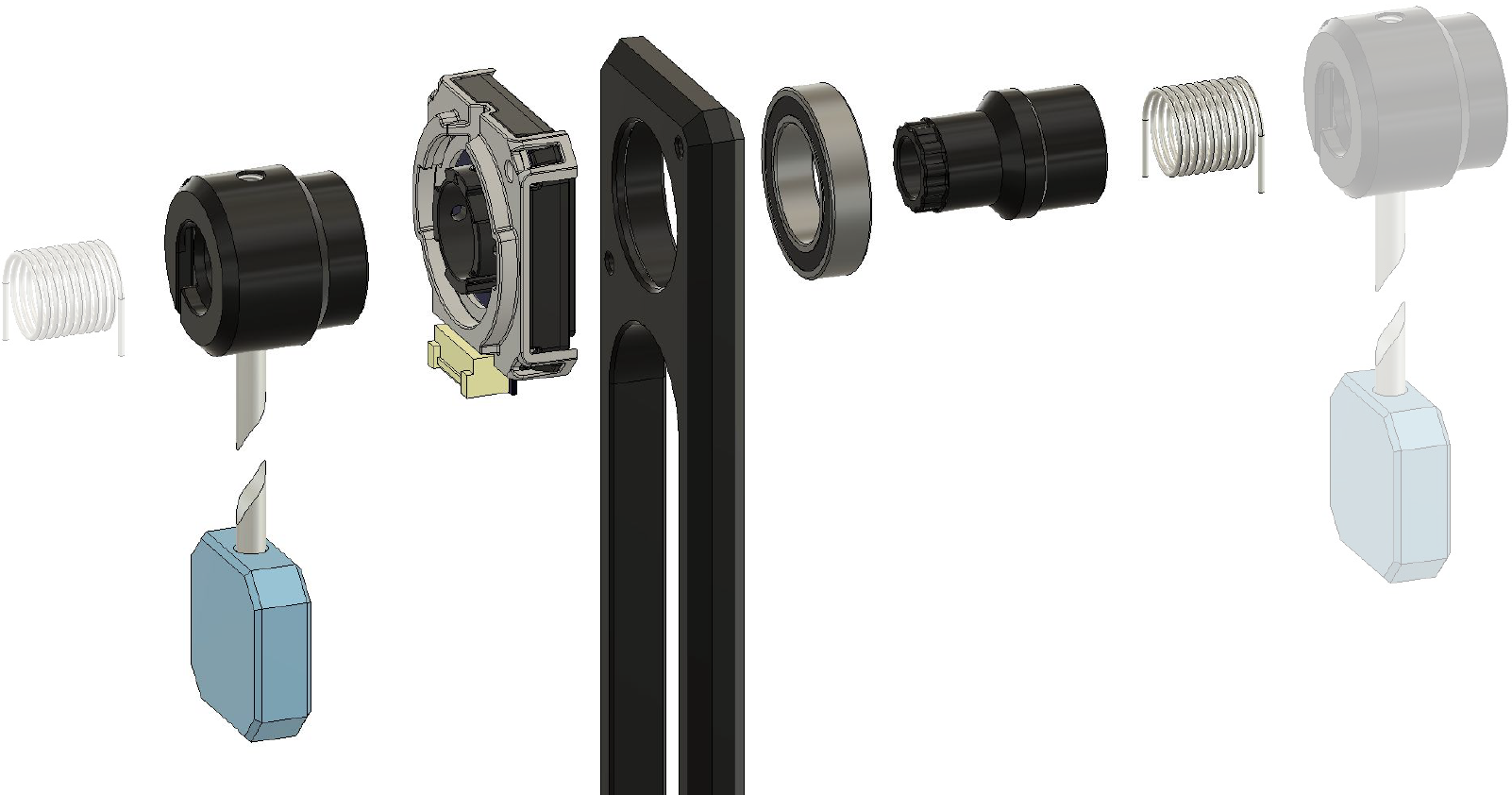}
        \caption{3D visualization of the module's mechanical design}\label{fig:temp_FK_unit_fusion}
\end{figure}

The photo of our platform is in Fig.~\ref{fig:temp_FK_lab_model_complete}.
The platform consists of 20 connected pendulum modules mounted on an aluminum profile, two motors attached at the boundaries, and electronics for control and data acquisition.
Each module, schematically visualized in Fig.~\ref{fig:temp_FK_unit_fusion}, comprises a frame with a pendulum, rotary encoder, and a torsional spring.
The mechanical parts of the module are designed as a combination of 3D-printed and off-the-shelf components.
Therefore, the module is adjustable and can be easily reproduced.
The total length of the platform with 20 modules is about $\SI{1}{\metre}$.
The approximate price of one module is €60.

\begin{figure}[!t]
        \centering
        \includegraphics[width=8.4cm]{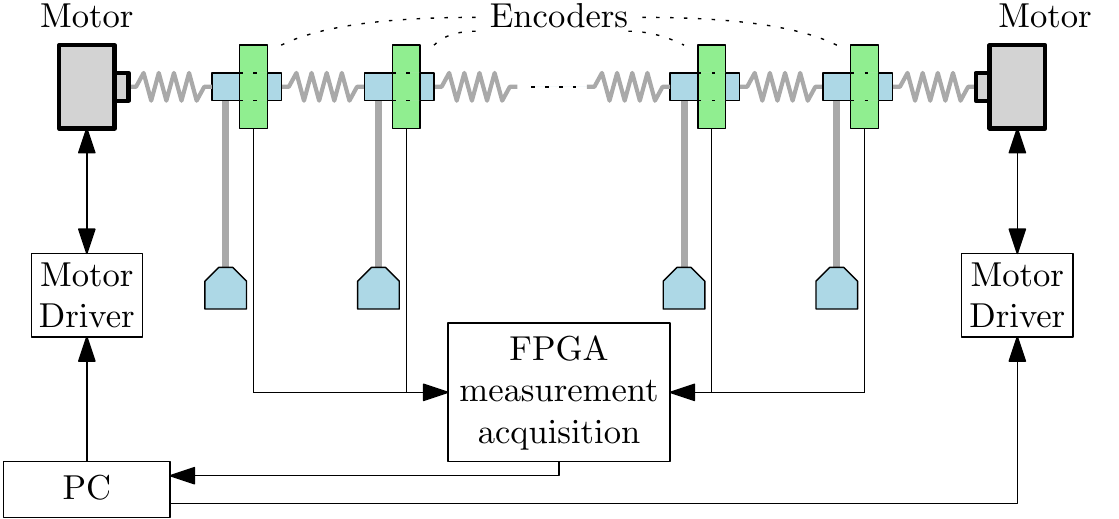}
        \caption{The hardware architecture of our platform}\label{fig:FK_HW_architecture}
\end{figure}

The pendulum's angle is measured using a capacitive encoder AMT132S-V by \textit{CUI Devices} with a resolution of~\num{4096} pulses per one revolution.
To process the measurements, we used FPGA development board DE0-nano by \textit{Terasic} with a custom-made shield (add-on), allowing read-out with a frequency up to $\SI{500}{\hertz}$.
The motors actuating the chain are NEMA17 $\SI{1.8}{\degree}$ stepper motors controlled by \textit{Pololu} Tic T249 drivers.
The data acquisition board and motor controllers are then connected to the PC that runs the control algorithms.
Fig.~\ref{fig:FK_HW_architecture} displays the HW architecture.

\begin{remark}  
        Stepper motors are able to directly and accurately set their desired position or speed. 
        Thus, we can view the stepper motors attached through the springs to the boundary pendulums as ideally controlled \textit{virtual} pendulums.
        However, different motors (e.g., BLDC motors) need to be used if one wants to control the exerted torque on boundary pendulums instead.
\end{remark}

\section{Mathematical Model}
We denote the $i$-th pendulum angle as $\varphi_i$.
The total torque acting on each pendulum can be decomposed into several terms.
For an uncoupled pendulum, the torques come from gravity and friction in the bearing. 
We model the friction as a simple linear dependence on a pendulum's absolute angular speed $\dot{\varphi}$ with a coefficient~$\gamma$. 

The nearest-neighbor coupling via real springs adds two additional torques. 
The first torque comes from potential energy accumulated in the spring, and the second torque results from energy dissipation in a real spring.
We model the dissipation again as friction linearly dependent on relative speeds of adjacent pendulums with a coefficient~$b$.
The equation describing a motion of $i$-th pendulum, $i = 1, \ldots, N$, is
\begin{equation}\label{eq:main_model} 
\begin{split}
J\ddot{\varphi_i}       &+ mgl\sin(\varphi_i) + \gamma \dot{\varphi_i} - \frac{k}{2}\pdv{\varphi_i}\sum_{i=1}^{N-1}(\varphi_{i+1} - \varphi_i)^2 \\
                        &- \frac{b}{2}\pdv{\dot{\varphi}_i} \sum_{i=1}^{N-1} (\dot{\varphi}_{i+1} - \dot{\varphi}_i)^2  =  M_i \;,
\end{split}
\end{equation}
where $g$ is the gravity constant, $k$ is spring constant, and $J$, $m$, and $l$ are the moment of inertia, mass, and length of the pendulum, respectively.
The terms $M_i$ represent external torques to be specified.

\begin{figure}[!t]
        \centering
        \includegraphics[width=8.4cm]{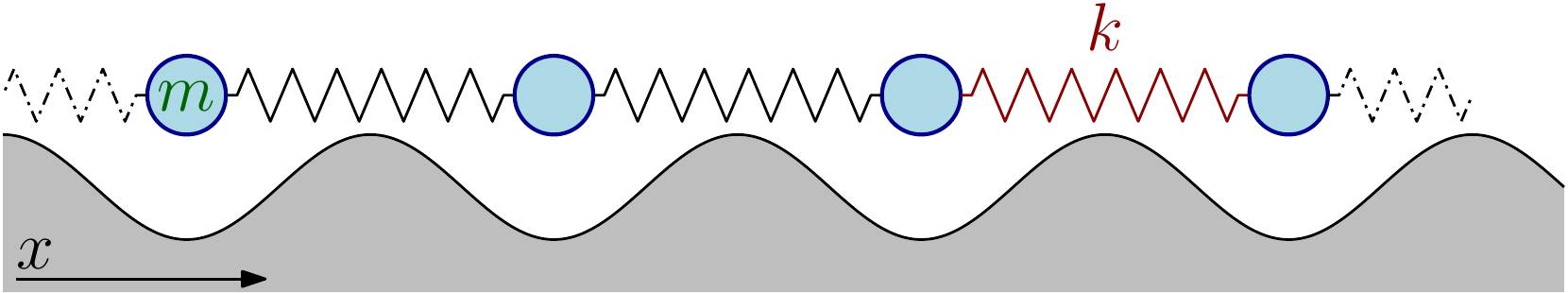}
        \caption{The mechanical analogy of the FK model as a chain of particles that slides over a potential landscape}\label{fig:FK_chain_mass}
\end{figure}

\begin{remark}
        Another mechanical analogy of the FK model is depicted in Fig.~\ref{fig:FK_chain_mass}, where the translational coordinate $x$ replaces the angular coordinate $\varphi$.
        This mechanical analogy was originally presented in~\cite{frenkel_theory_1938}.
\end{remark}

\begin{remark}
        The standard FK model proposed in~\cite{frenkel_theory_1938} differs from the presented model~\eqref{eq:main_model} in two aspects.
        First, the standard model is defined without dissipation terms, and second, with ${N \rightarrow \infty}$ and, thus, no boundaries.
\end{remark}


\subsection{External Inputs}
The laboratory platform can be actuated by two motors attached through springs to the boundary pendulums, schematically depicted in Fig.~\ref{fig:FK_HW_architecture}.
To specify the torques $M_i$ in~\eqref{eq:main_model}, we denote the angles of motors attached to the first and the last pendulum in the chain as $\varphi_\mathrm{M_1}$, and $\varphi_\mathrm{M_2}$, respectively.
The external input torques can be written as
\begin{equation} 
        M_i = 
        \begin{cases}
                k (\varphi_\mathrm{M_1} - \varphi_1) + b (\dot{\varphi}_\mathrm{M_1} - \dot{\varphi}_1)         &       \text{for } i=1  \;, \\
                k (\varphi_\mathrm{M_2} - \varphi_N) + b (\dot{\varphi}_\mathrm{M_2} - \dot{\varphi}_N)         &       \text{for } i=N  \;,\\
                0       &       \text{otherwise}        \;.
        \end{cases}
\end{equation}

\subsection{Equations in a Matrix Form}
To write the model~\eqref{eq:main_model} in a state-space form, we use the formalism of multiagent systems.
The topological structure of the interactions between the pendulums can be described by an undirected path graph with a graph Laplacian ${L = (l_{ij}) \in \mathbb{R}^{N \times N}}$:
\begin{equation}\label{eq:Laplacian}
        L = 
        \begin{bmatrix}
        1     & -1    & 0     &     &\cdots& &  0     \\
        -1    & 2     & -1    &     &\cdots& &  0     \\
        0     &-1     & 2     &     &\cdots& &  0     \\
        \vdots&       &  & \ddots    &     &  &  \vdots\\
        0     &       &\cdots& &2    & -1    &  0     \\
        0     &       &\cdots& &-1   & 2     &  -1     \\
        0     &       &\cdots& &0    & -1    &  1   
        \end{bmatrix}\;.
\end{equation}
Furthermore, by defining the state of the $i$-th pendulum as ${x_i = [\varphi_i, \dot{\varphi}_i]\tran = [x_{i,1}, x_{i,2}]\tran}$, the state-space equations describing the dynamics can be written as
\begin{equation}\label{eq:FK_model_MAS} 
  \dot{x}_i = f\left(x_i\right) - \frac{1}{J} B \left( K \sum_{j = 1}^N l_{ij} x_j - M_i \right)\;,
\end{equation}
where $f(x_i)$ is the uncoupled dynamics of a single pendulum
\begin{equation}\label{eq:drift_dynamics_MAS} 
  f(x_i) =
  \begin{bmatrix}
    x_{i,2} \\
    -\dfrac{mgl}{J}\sin(x_{i,1}) -\dfrac{\gamma}{J} x_{i,2} 
  \end{bmatrix}\;,
\end{equation}
and ${B = [0, 1]\tran}$, ${K = [k, b]}$, which render the torque coupling through springs with dissipation.
Additionally, let 
\begin{equation} 
        \begin{split}
                x &= [x_1\tran, \ldots, x_N\tran]\tran           \;,    \\
                u &= [\varphi_\mathrm{M_1}, \dot{\varphi}_\mathrm{M_1}, \varphi_\mathrm{M_2}, \dot{\varphi}_\mathrm{M_2}]\tran \;, \\
                F(x) &= [f(x_1)\tran, \ldots, f(x_N)\tran]\tran  \;,    \\
                d_\mathrm{M_1} &= [1, 0, \ldots, 0]\tran \in  \mathbb{R}^{N} \;, \\
                d_\mathrm{M_2} &= [0, \ldots, 0, 1]\tran \in  \mathbb{R}^{N} \;, \\
                D &= \mathrm{diag}(d_\mathrm{M_1} + d_\mathrm{M_2})             \;.   \\
        \end{split}
\end{equation}
The final compact matrix form of the system with inputs is
\begin{equation}\label{eq:matrix_form_model_main} 
        \begin{split}
                \dot{x} =& F(x) - \left( \frac{1}{J} (L+D) \otimes B K \right) x  \\
                         &+ \left( \frac{1}{J} \left[ d_\mathrm{M_1}, d_\mathrm{M_2} \right] \otimes B K \right) u \;,
        \end{split}
\end{equation}
where $\otimes$ denotes the Kronecker product.


\subsection{Parameter Identification and Model Verification}\label{sec:identf_verif}

\begin{table}[!tb] 
\begin{center}
\caption{Mechanical parameters of the platform}\label{tab:platform_parameters}
\begin{tabular}{lcc}
Description                     & Symbol        & Value \\\hline
Rod length                      & $l$           & $\SI{0.15}{\metre}$\\
Nominal spring constant         & $\bar{k}$& \SI{0.054}{\newton\per\metre}  \\
Pendulum's weight               & $m$           & $\SI{17}{\gram}$ \\
Moment of inertia               & $J$           & $\SI{3.82e-4}{\kilo\gram\metre^2}$  \\
\hline
Identif. spring constant        & $k$      & $\SI{0.065}{\newton\per\metre}$ \\
Relative dissipation coef.         & $b$           & $\SI{1.70e-3}{\newton\meter\second\per\radian}$  \\ 
Absolute dissipation coef.         & $\gamma$      & $\SI{3.75e-4}{\newton\meter\second\per\radian}$
\end{tabular}
\end{center}
\end{table}

To verify the validity of the presented mathematical model, we first identify the unknown parameters.
As known parameters, we choose the mass $m$ and length $l$, from which we also computed the moment of inertia $J=ml^2$, assuming negligible mass of the rod.
The numerical values are listed in the first part of Tab.~\ref{tab:platform_parameters}.
The parameters we need to identify are the spring constant $k$ and the two damping coefficients, $\gamma$ and $b$.
Although the nominal spring constant $\bar{k}$ is known from the datasheet, the value is valid only when the spring is rotated along its wind direction. 
Since the spring can rotate in both directions in our case, we choose $k$ to be unknown.

\begin{figure}[!t]
        \centering
        \includegraphics[width=8.4cm]{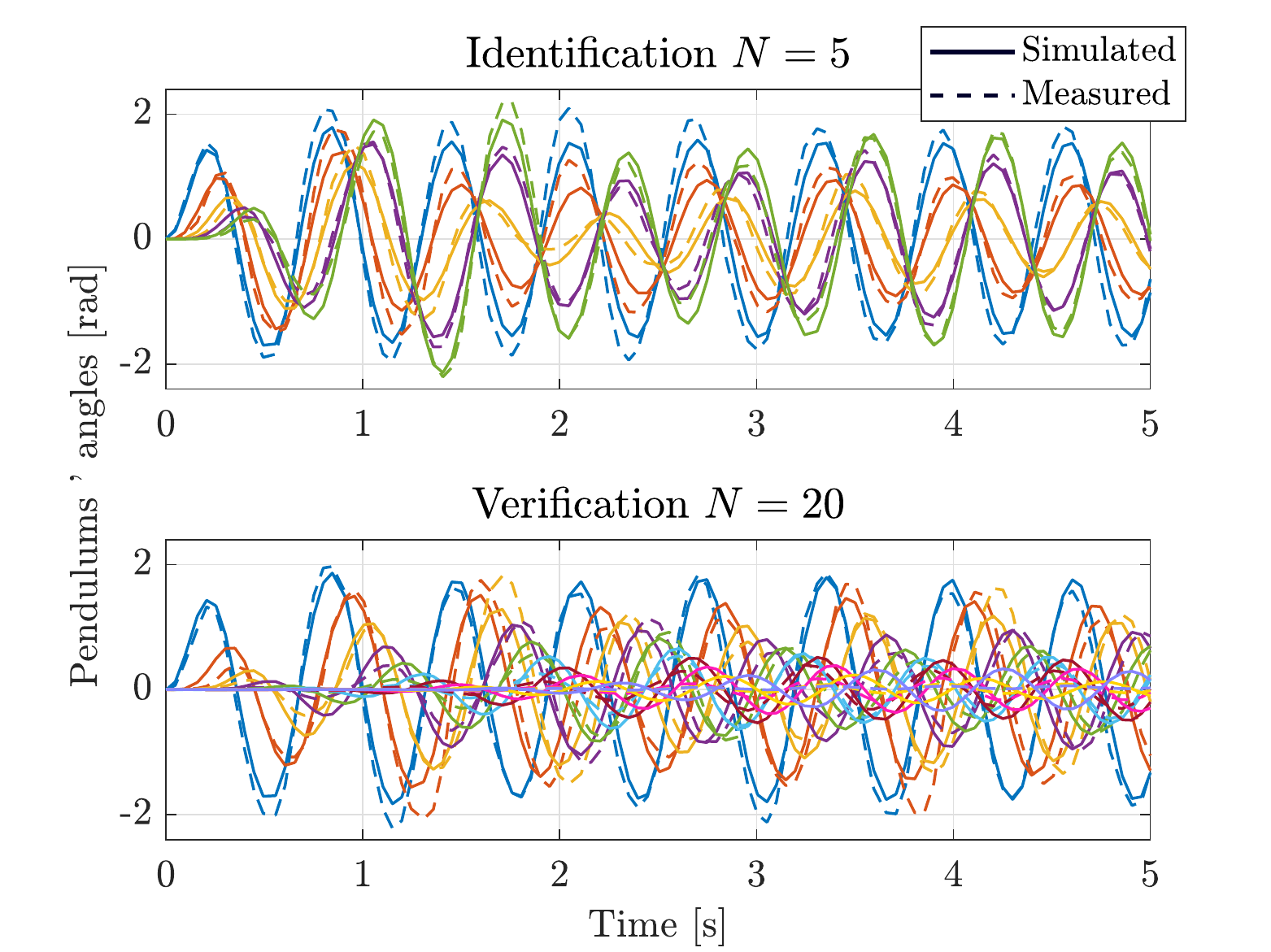}
        \caption{Simulation output with identified parameters compared with measurements from the real platform.
                Identification and verification were done with $N=5$ and $N = 20$ pendulums, respectively.}\label{fig:identf_verif}
\end{figure}

The identification was defined as a gray box model, and the unknown parameters were found using System Identification Toolbox for Matlab, minimizing the least square error of all pendulum's angles in the chain.
To collect input data for the identification, we used a chain of $N=5$ pendulums with only one motor's torque $M_{1}$ active ($M_N = 0$) and set the motor's trajectory to
\begin{equation}\label{eq:verif_input} 
        \varphi_\mathrm{M_1}(t) = a \sin(\omega t) \;,
\end{equation}
where $a$ and $\omega$ are the signal's amplitude and the angular frequency, respectively.
Several values of $a$ and $\omega$ were used to collect many identification trajectories.
For evaluation, we used the signal~\eqref{eq:verif_input} with $a = \SI{2}{\radian}$ and $\omega = \SI{10}{\radian\per\second}$.
See the top part of Fig.~\ref{fig:identf_verif} comparing measured pendulums' angles and outputs from simulating the model~\eqref{eq:matrix_form_model_main} with identified parameters.
In the identification experiment, the normalized root-mean-square error (NRMSE) averaged over all pendulums in the chain was $e_\text{idn} = 0.12$.

For the verification, we compare the real system and the output from the simulation with $N = 20$ pendulums actuated with the same input~\eqref{eq:verif_input}, see the bottom part of Fig.~\ref{fig:identf_verif}.
The NRMSE averaged over all pendulums in the verification experiment was $e_\text{vrf} = 0.28$.
One can see that the error $e_\text{ver}$ grows with an increasing number of pendulums in the chain. 
However, the main behavior of the system is well captured by the model.

\section{Examples of Boundary Control Formulations}\label{sec:examples_ctrl}
\begin{figure}
        \begin{subfigure}[t]{0.25\textwidth}
                \centering
                \includegraphics[height=3cm]{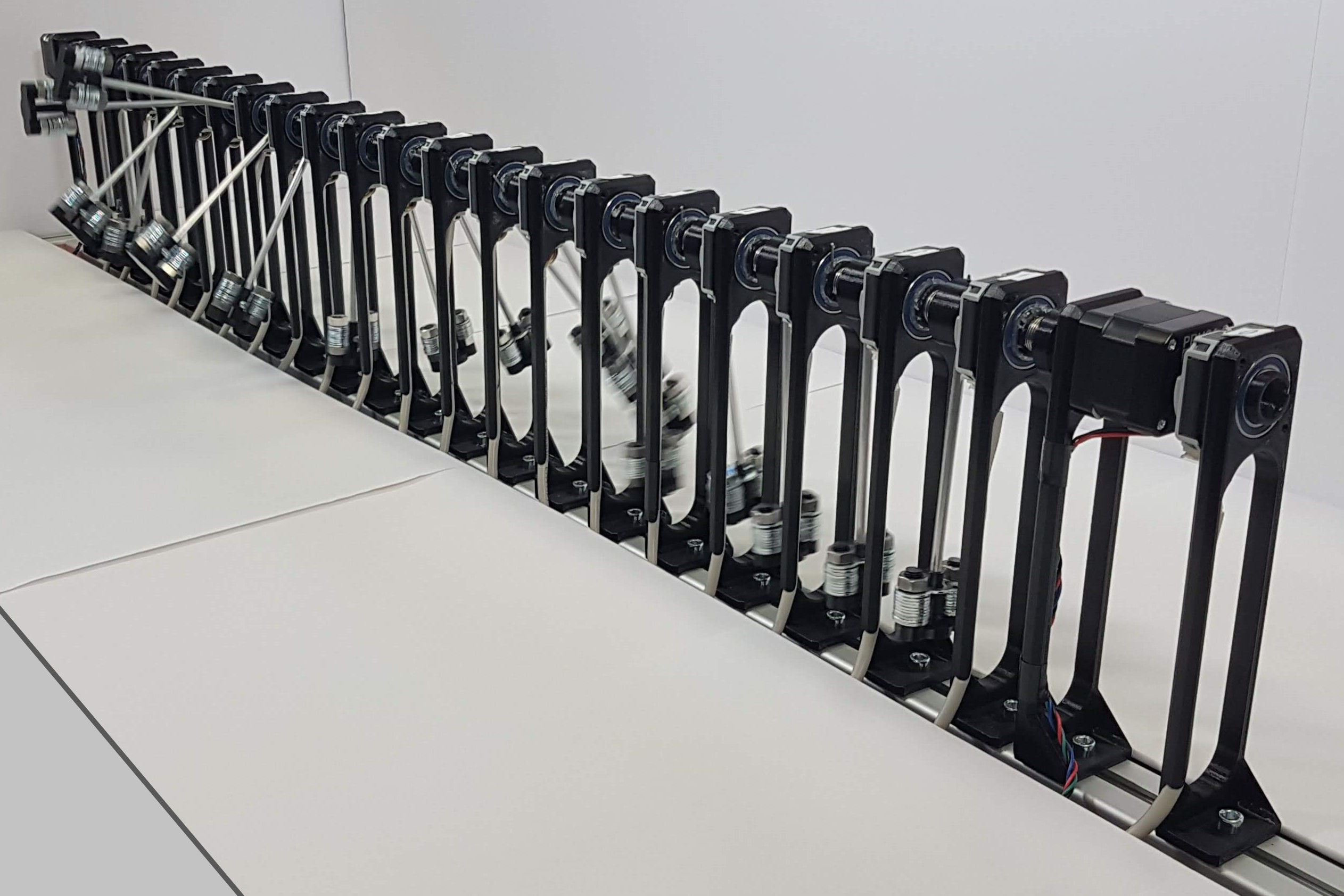}
                \caption{}
        \end{subfigure}
        \hfill
        \begin{subfigure}[t]{0.15\textwidth}
                \centering
                \includegraphics[height=3cm]{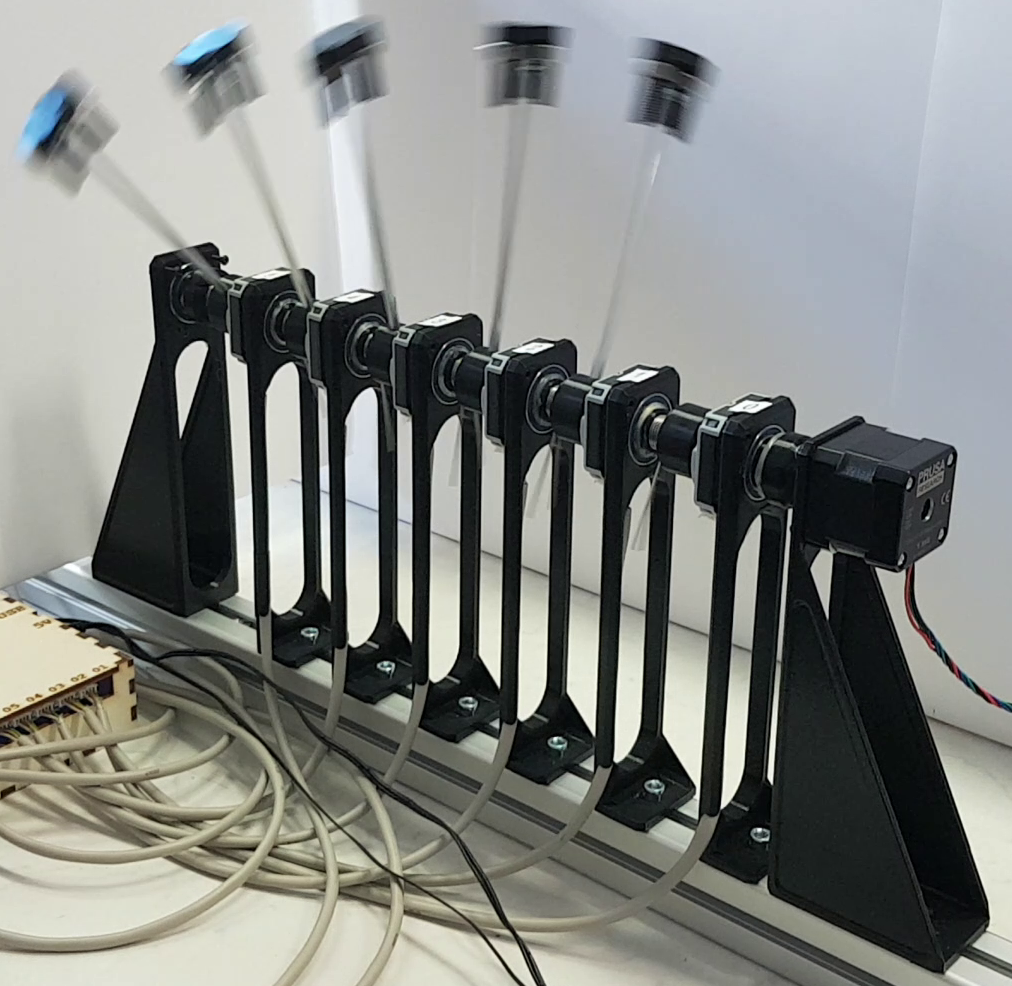}
                \caption{}
        \end{subfigure}
        \caption{Snapshots from the experiments: (a) Non-collocated control of oscillations, (b) Open-loop, low-oscillatory rotation of the pendulums}\label{fig:photos_experiments}
\end{figure}

As described in previous sections, the system is boundary-controlled.
Thus, for $N > 2$, the system is underactuated, which imposes challenging control problems. 
This section proposes two such control formulations and describes control algorithms for solving them.
We first designed and tuned the algorithms on the mathematical model, and then the solutions were verified on the real platform.
The snapshots from the two experiments are in Fig.~\ref{fig:photos_experiments}.

\subsection{Non-collocated Control of Oscillation}
As the first task, we propose feedback control of a single pendulum's oscillations.
Consider a pendulum in the chain with an index ${i^\star > 1}$ that we select to be stabilized to ${\varphi^\star = 0}$ while the torque $M_1(t)$ is controlled and $M_N(t)$ acts as an unknown disturbance.
We can formulate the goal to satisfy a condition
\begin{equation}\label{eq:formulation_noncollocated}
        \lim_{t \rightarrow \infty} | \varphi_{i^\star}(t) - \varphi^\star | \leq \epsilon \;,
\end{equation}
where $\epsilon \geq 0$ is some small constant.
The challenge is that the input $M_1(t)$ is \textit{non-collocated} to the $i^\star$-th pendulum, with the coupling represented by the \textit{inter}-pendulums, which is non-linear.

To solve problem~\eqref{eq:formulation_noncollocated}, we view the action of the disturbance $M_N(t)$  as launching a traveling wave into the system. 
The idea is to launch, using $M_1(t)$, a traveling wave that reaches the $i^\star$-th pendulum with the same amplitude as the disturbance wave but with the opposite phase.
The amplitudes of the two waves cancel out at position index~$i^\star$, so ${|\varphi_{i^\star}(t)| \rightarrow 0}$.
This concept of wave-based control of oscillations in lump-modeled flexible structures was used in many works; for instance, see~\cite{oconnor_wave-based_2007} and references therein.


\subsubsection{Naive Solution}
With available measurements from all pendulums and if $i^\star \leq (\frac{N}{2} + \Delta)$, where $\Delta$ is some small constant integer (see Remark~\ref{rem:solution_wave}), the solution is straightforward.
The control law for the controlled motor to launch a wave that cancels the disturbance at index $i^\star$ is
\begin{equation}\label{eq:naive_solution_wave} 
        \varphi_\mathrm{M_1}(t) = - \varphi_{2i^\star}(t) \;.
\end{equation}  
The validity of control law~\eqref{eq:naive_solution_wave} to satisfy goal~\eqref{eq:formulation_noncollocated} can be derived from the system's symmetry, which is evident from the Laplacian~\eqref{eq:Laplacian} of the system.
Thus, by applying~\eqref{eq:naive_solution_wave}, the net torque acting on the $i^\star$-th pendulum is zero after transient effects. 



\subsubsection{Wave-based Delay Compensation}
However, on a real platform, the solution~\eqref{eq:naive_solution_wave} is not directly applicable.
The main reason is the inevitable time delay $t_\mathrm{d}$ in the feedback loop caused by the hardware implementation.
So instead of the measurement $\varphi_{2i^\star}(t)$ being available at the time $t$, only $\varphi_{2i^\star}(t - t_d)$ is available.
Using the delayed measurement in~\eqref{eq:naive_solution_wave} would naturally cause phase mismatch of the waves reaching the $i^\star$-th pendulum.
Thus, the goal~\eqref{eq:formulation_noncollocated} would not be guaranteed.
There are many ways how one could compensate for the delay, for instance, using the \textit{Smith predictor} or possibly the \textit{Kalman Filter} to estimate $\varphi_{2i^\star}(t)$ from delayed measurements.
The system is, however, non-linear.

\begin{figure}[!tb]
        \centering
        \includegraphics[width=8.4cm]{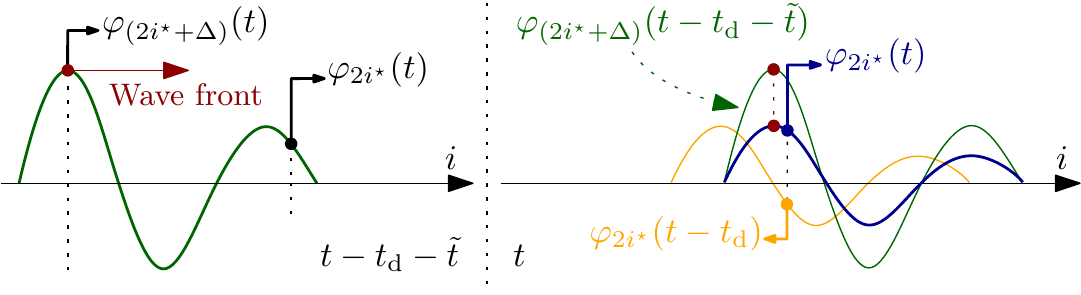}
        \caption{Wave-based compensation of the feedback delay $t_\mathrm{d}$.}\label{fig:wave_based_delay_compensation_schemeV2}
\end{figure}

Here, we propose yet another way to compensate for the delay using the wave character of the system.
The solution is schematically depicted in Fig.~\ref{fig:wave_based_delay_compensation_schemeV2}.
Consider a particular wave front traveling through the system in the direction of decreasing position index, and let define $t(i, j)$ to be the time needed for a wave front to travel from $i$-th to $j$-th pendulum.
As the wave travels through the system, it follows from the wave property of the system that each pendulum delays the wave front by $t(i+1, i)$ and decreases the wave front's amplitude by some coefficient $\lambda$.
Therefore, to compensate for the phase mismatch caused by the delay $t_\mathrm{d}$, we can use, instead of $\varphi_{2i^\star}(t)$, a measurement $\varphi_{(2i^\star + \Delta)}(t - t_d - \tilde{t})$, where $\Delta = 1, 2, \ldots$ and ${\tilde{t} \geq 0}$ are selected to satisfy
\begin{equation}
        t(2i^\star+\Delta, 2i^\star) - \tilde{t} = t_\mathrm{d} \;,
\end{equation}
and $\tilde{t} < t(i+1,i)$.
The control law compensating the time delay is
\begin{equation}\label{eq:WBC_solution_delay_compensation}
        \varphi_\mathrm{M_1}(t) = - \lambda \varphi_{(2i^\star+\Delta)}(t - t_\mathrm{d} - \tilde{t}) \;, 
\end{equation}  
where $\lambda \geq 0$ is a parameter to compensate for a gain mismatch of the wave fronts.
We describe the method to obtain the parameter $\lambda$ below.



\subsubsection{Adaptive Gain Compensation}
\begin{figure}[!t]
        \centering
        \includegraphics[width=8.4cm]{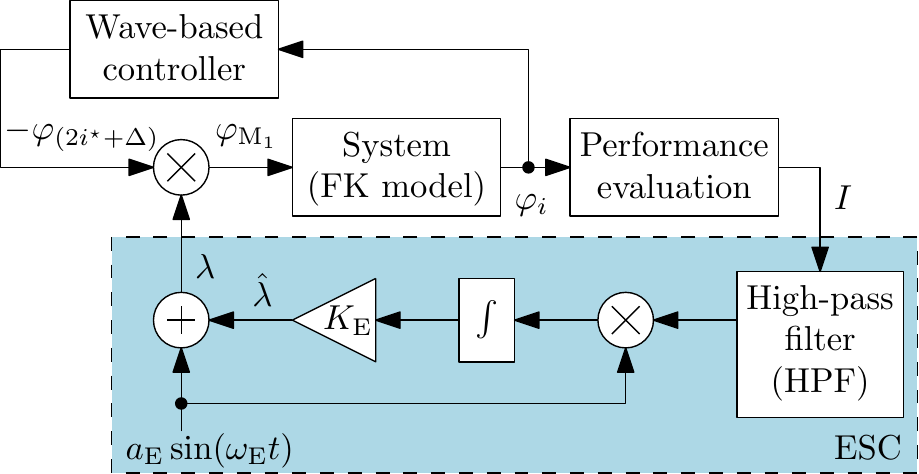}
        \caption{Block diagram of wave-based non-collocated control of oscillations with an extremum seeking controller adaptively changing the control gain}\label{fig:ESC_diagram}
\end{figure}

To find the gain $\lambda$, we employ the \textit{Extremum Seeking Control} (ESC); see~\cite{ariyur_real-time_2003} for details.
The ESC allows to find the optimal value of $\lambda$ based on online minimization of a given performance index $I$. 
Since the goal is to stabilize the pendulum at ${\varphi^\star = 0}$, we chose the performance index to be minimized as a running average of the selected pendulum's angle $\varphi_{i^\star}$ in an absolute value, that is
\begin{equation}\label{eq:performace_index} 
        I(t) = \frac{1}{W} \sum_{j = 0}^{W-1} | \varphi_{i^\star}(t-jT_\mathrm{s}) | \;,
\end{equation}
where $W$ is the time window size over which the running average is computed.
The block diagram of the ESC implementation is depicted in Fig.~\ref{fig:ESC_diagram}.

\begin{remark}\label{rem:solution_wave}
        The formulated solution to the problem~\eqref{eq:formulation_noncollocated} works only when the index of the selected pendulum is ${i^\star \leq (\frac{N}{2} + \Delta)}$.
        Naturally, for ${i^\star> (\frac{N}{2} + \Delta)}$, the measurement ${\varphi_{(2i^\star + \Delta)}}$ is not available to implement~\eqref{eq:WBC_solution_delay_compensation}.
        A different solution needs to be designed for such cases.
\end{remark}


\subsubsection{Experiment}

\begin{table}[!tb]
        \begin{center}
        \caption{Non-collocated control: experiment parameters}\label{tab:non_collocated_exp_param}
        \begin{tabular}{lcc}
        Description                     & Symbol                & Value \\\hline
        Control and sampling period     & $T_\mathrm{s}$        & \SI{0.03}{\second}    \\
        Disturbance: amplitude          & $a_\mathrm{M_2}$       & \SI{3}{\radian}       \\
        Disturbance: angular frequency  & $\omega_\mathrm{M_2}$  & \SI{9.24}{\radian\per\second}     \\
        \hline
        ESC: Size of the time window    & $W$                   & 20     \\
        ESC: Gain                       & $K_\mathrm{E}$        & 8                     \\
        ESC: Sine frequency             & $\omega_\mathrm{E}$   & \SI{0.5}{\hertz}      \\
        ESC: Sine amplitude             & $a_\mathrm{E}$        & 0.01                  \\
        ESC: HPF cut-off frequency      & $f_\mathrm{HPF}$      & \SI{0.1}{\hertz}
        \end{tabular}
        \end{center}
\end{table}

\begin{figure}[!t]
        \centering
        \includegraphics[width=8.4cm]{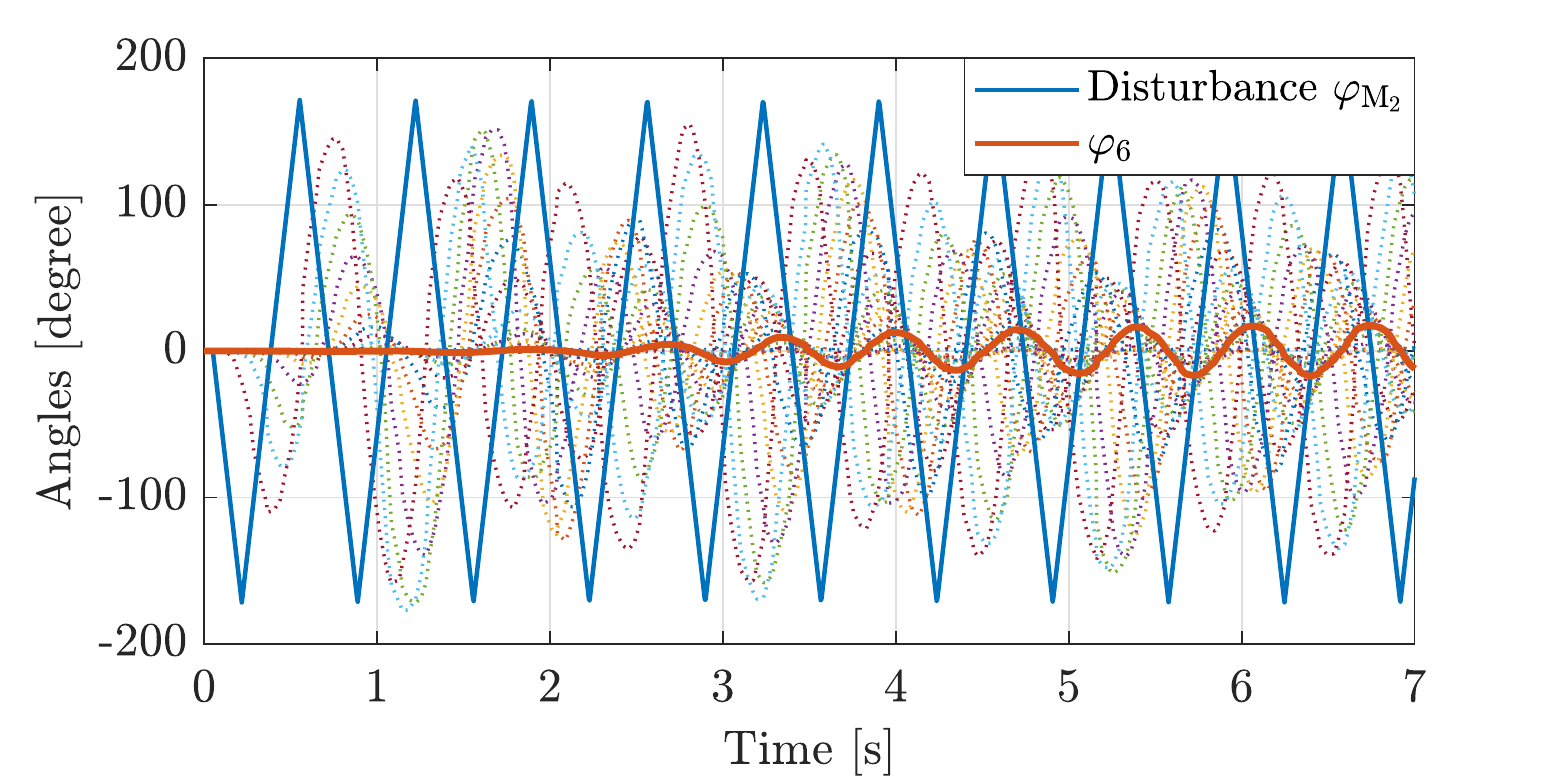}
        \caption{Response of the system to the disturbance~\eqref{eq:disturbance}. Highlighted is the disturbance signal and the angle of the pendulum which we wish to stabilize.}\label{fig:response_disturb_ESC}
\end{figure}

To experimentally verify and demonstrate the designed control, we used $N=20$ pendulums and selected the pendulum $i^\star=6$ to be stabilized.
We set the disturbance's trajectory as a periodic triangle wave signal
\begin{equation}\label{eq:disturbance} 
        \varphi_\mathrm{M_2}(t) = -\frac{2a_\mathrm{M_2}}{\pi} \arcsin \left(  \sin \left(  \omega_\mathrm{M_2} t \right)  \right) \;.
\end{equation}
The system's response to the disturbance is shown in Fig.~\ref{fig:response_disturb_ESC}.
Numerical values of the parameters related to the experiment are in Tab.~\ref{tab:non_collocated_exp_param}.
The parameters of the ESC were selected to achieve the best performance.

\begin{figure}[!t]
        \centering
        \includegraphics[width=8.4cm]{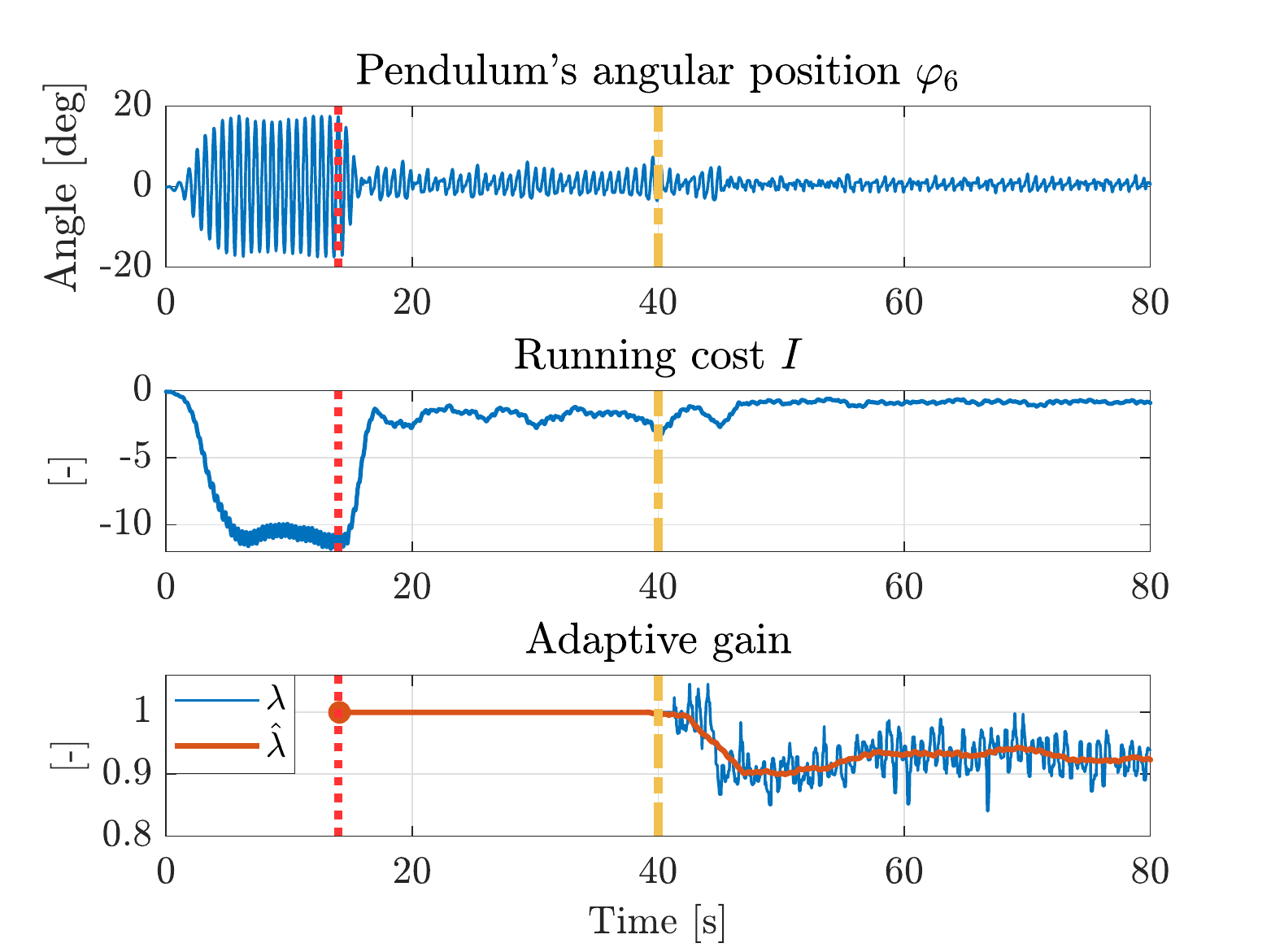}
        \caption{Experimental verification of the proposed non-collocated, wave-based control of single pendulum's oscillation}\label{fig:wave_esc_P06_experiment}
\end{figure}

We divide the experiment into three time intervals to showcase the designed control; see Fig.~\ref{fig:wave_esc_P06_experiment}.
In the first interval starting at $t = \SI{0}{\second}$, only the disturbance $M_N$ is activated while the control loop is not.
The maximum amplitude of the selected pendulum is $\varphi_{6, \max} \approx \SI{18}{\degree}$. 
At $t = \SI{14}{\second}$, the solution \eqref{eq:WBC_solution_delay_compensation} with $\lambda = 1$ is activated, resulting in a decrease of the oscillation amplitude to $\varphi_{6, \max} \approx \SI{5}{\degree}$.
Lastly, in the third interval, the adaptive loop with ESC is turned on, and after $\approx \SI{10}{\second}$, it additionally decreases the maximal amplitude to $\varphi_{6, \max} \approx \SI{2}{\degree}$.
The remaining oscillations are mainly due to the persisting ESC, changing the gain $\lambda$.

\subsection{Open-loop, Low-oscillatory Rotation of the Pendulums}
The second control problem we demonstrate is the following.
Consider the system~\eqref{eq:main_model} controlled only by one motor $M_1(t)$.
Given a reference speed $\omega_\mathrm{ref} \neq 0$, the goal is to rotate all pendulums in the chain with an average speed equal to $\omega_\mathrm{ref}$ while minimizing the oscillations within the chain. 
Specifically, we define the goal as to find the reference trajectory $x_0(t)$ for which
\begin{subequations}\label{eq:low_oscillatory_control_goal} 
        \begin{align}
                \min & \sum_{i > j} \int_0^{T_\mathrm{f}} | x_{i,2}(t) - x_{j,2}(t) |  \mathrm{d}t  \;,  \label{eq:low_oscillatory_control_goal_A} \\
                \lim_{t \rightarrow \infty} & \bar{x}_{i,2}(t)  = \omega_\mathrm{ref} \;, i = 1, 2, \ldots, N  \;, \label{eq:low_oscillatory_control_goal_B}
        \end{align}
\end{subequations}
where $\bar{(.)}$ denotes the mean value and $T_\mathrm{f}$ is the final time.
As we argued in~\cite{do_synchronization_2021}, the formulated problem~\eqref{eq:low_oscillatory_control_goal} might be relevant for the control of nanoscale friction.

\subsubsection{Synchronization-based Solution}
We can address the goal~\eqref{eq:low_oscillatory_control_goal_A} as to reach synchronization in the pendulums' speeds.
Consider trajectory $x_0(t)$ to be a solution to the drift dynamics of a single pendulum~\eqref{eq:drift_dynamics_MAS}.
We define the synchronization error ${\delta_i = x_i - x_0}$, which gives the error dynamics
\begin{equation}\label{eq:error_dynamics}
        \dot{\delta}_i = \dot{x}_i - \dot{x}_0 = f(x_i) - f(x_0)  - \frac{1}{J} B \left( K \sum_{j = 1}^N l_{ij} x_j + M_i \right)\;,
\end{equation}
with $M_1(t) = K(x_1 - x_0)$.
By defining $\delta = [\delta_1\tran, \ldots, \delta_N\tran]\tran$, we can rewrite~\eqref{eq:error_dynamics} into a matrix form
\begin{equation}\label{eq:error_dynamics_matrix_form}
        \dot{\delta} = \mathcal{F}(\delta, x_0) = F(x) - F(x_0) - \left( (L+D)	\otimes (BK) \right) \delta \;,
\end{equation}
Linearizing the system~\eqref{eq:error_dynamics_matrix_form} around the origin yields the Jacobian matrix
\begin{equation}\label{eq:linear_error_dyn}
          \left.
          \pdv{\mathcal{F}(\delta, x_0)}{\delta}\right\vert_{\begin{subarray}{l} \delta =0 \\ x_0 = 0\end{subarray}}
          =
            I_N \otimes 
              \begin{bmatrix}
                0  & 1 \\
                -1 & -\gamma
              \end{bmatrix}
              -
              (L+D) \otimes 
                BK
                \;,
\end{equation}
where $I_N$ is an identity matrix of size $N$.
By examining the eigenvalues of the matrix~\eqref{eq:linear_error_dyn}, it can be shown that the origin of the system~\eqref{eq:error_dynamics_matrix_form} is locally asymptotically stable; see~\cite{do_synchronization_2021} for a more detailed analysis.
Thus, we have shown, that pendulums completely synchronize by choosing $x_0(t)$ as a solution to~\eqref{eq:drift_dynamics_MAS}.
However, the solution of~\eqref{eq:drift_dynamics_MAS} with $\gamma > 0$ corresponds to a trajectory of a single, damped pendulum.
Such trajectory asymptotically decays to the origin with $x_{i, 2} \rightarrow 0$, so the second goal~\eqref{eq:low_oscillatory_control_goal_B} could not be satisfied.

To satisfy the second condition, we consider the $\tilde{x}_0(t)$ to be a solution to~\eqref{eq:drift_dynamics_MAS} with $\gamma = 0$ and such initial conditions that the pendulum has enough energy to swing through the inverse position.
This results in a continuous rotation with $\bar{x}_{i,2} > 0$ while the pendulums are able to move close to synchrony as the absolute dissipation in the real system acts only as a perturbation to the complete synchrony.
The proposed solution to the goal~\eqref{eq:low_oscillatory_control_goal_B} is open-loop; the motor tracks the given fixed trajectory $\tilde{x}_0$.

\subsubsection{Experiment}
\begin{figure}[!t]
        \centering
        \includegraphics[width=8.4cm]{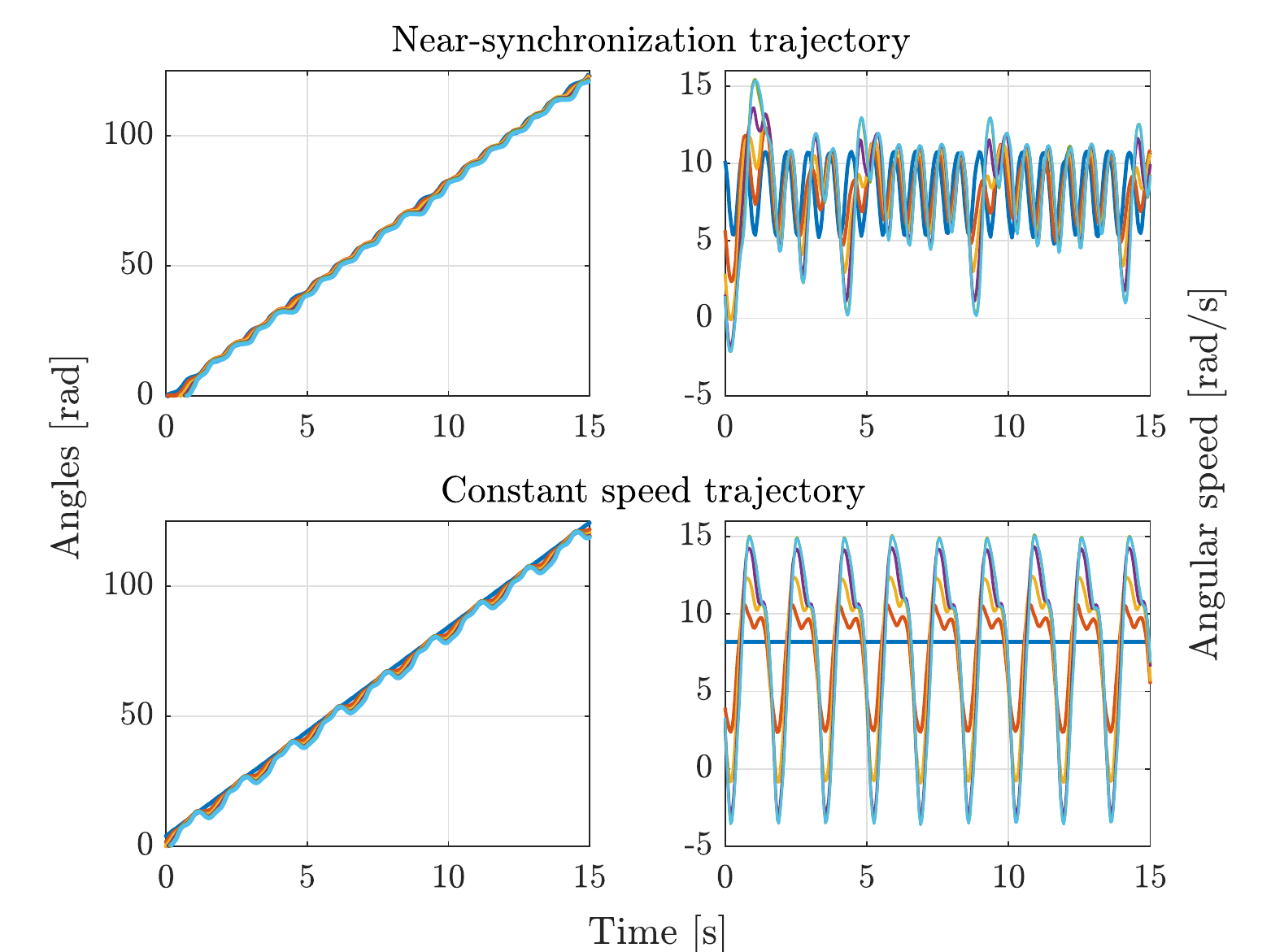}
        \caption{Experiments: comparison of near-synchronization reference trajectory and constant speed trajectory. 
        The near-synchronized trajectory allows to considerably reduce the oscillations within the chain.}\label{fig:comparison_sync_const_speed}
        % ... 
\end{figure}

We used $N = 5$ pendulums for the experiments as the coupling strength between the pendulums was not strong enough to keep more pendulums near synchrony.  
We set the average reference speed to ${\omega_\mathrm{ref} = \SI{8.2}{\radian\per\second}}$. 
To obtain the near-synchronized trajectory with a given speed, we set the initial conditions in~\eqref{eq:drift_dynamics_MAS} to $x_0(0) = [\pi, 3]\tran$ and $\gamma = 0$.
The results from the experiment are in the top part of Fig.~\ref{fig:comparison_sync_const_speed}.
Although the near-synchronized motion is being disrupted, the pendulums are able to re-synchronize and maintain the near-synchronized state for several periods.
The perturbations are mainly caused by accumulated mismatch between the real dynamics with $\gamma > 0$ and the reference trajectory.

To show the significance of the synchronized motion, we also run an experiment with a constant speed reference trajectory $x_0(t) = [\omega_\mathrm{ref}t, \omega_\mathrm{ref}]\tran$; see bottom part of Fig.~\ref{fig:comparison_sync_const_speed}.
We can see that as the pendulums are not synchronized, the oscillations are more significant, resulting in higher difference in pendulums' speeds. 
The criterion~\eqref{eq:low_oscillatory_control_goal_A} evaluated for the near-synchronized and the constant speed trajectories in the interval $t \in (0, 15)$ is for the former lower by $\SI{56}{\percent}$.

\section{Conclusion}
In this paper, we described our new laboratory mechatronic platform for experiments with a boundary control of an array/chain of some twenty coupled pendulums pivoting around a single shaft -- the mechanical realization of the Frenkel-Kontorova (FK) model. 
We also provided a mathematical model suitable for simulation experiments. 
Finally, we formulated two control problems, proposed methods for their solution, and demonstrated these experimentally using the platform. 
We offered the design files and source codes to the research community for free through a public repository. 
Our future work will address some admitted shortcomings of the solutions in this paper.

\bibliographystyle{IEEEtran}
\bibliography{biblio_files/biblio}

\end{document}